\documentclass{llncs}

\usepackage{amssymb,amsmath,amscd,latexsym}
\usepackage{algorithm}
\usepackage{luca}
\usepackage{url}
\usepackage{pstricks}
\usepackage{pst-node}
\usepackage{listings}
\usepackage{graphicx}
\usepackage{gastex}
\usepackage{subfigure}
\usepackage{paralist}

\def\qed{\rule{0.4em}{1.4ex}}

\newcommand{\PA}{0}
\newcommand{\PB}{1}

\newcommand{\SA}{S_0}

\newcommand{\SB}{S_1}

\newcommand{\SR}{S_{P}}

\newcommand{\gamegraph}{G}

\newcommand{\set}[1]{\{\: #1 \:\}}

\newcommand{\trans}{\delta}
\newcommand{\distr}{{\cal D}}

\newcommand{\Gist}{{\sc Gist}}
\newcommand{\GistUrl}{\url{http://pub.ist.ac.at/gist}}

\newcommand{\slopefrac}[2]{\leavevmode\kern.1em
  \raise .5ex\hbox{\the\scriptfont0 #1}\kern-.1em
  /\kern-.15em\lower .25ex\hbox{\the\scriptfont0 #2}}
\newcommand{\half}{\slopefrac{1}{2}}

\begin{document}
\title{\Gist: A Solver for Probabilistic Games}
\author{Krishnendu Chatterjee$^1$ \and Thomas A. Henzinger$^{1}$ \and
  Barbara Jobstmann$^2$ \and Arjun Radhakrishna$^1$}

\institute{IST Austria (Institute of Science and Technology Austria)
\and CNRS/Verimag, France}

\maketitle

\newif
  \iflong
  \longfalse
\newif
  \ifshort
  \shorttrue

\thispagestyle{empty}

\let\thefootnote\relax\footnotetext{This research was supported by the European Union project
COMBEST and the European Network of Excellence ArtistDesign.}

\begin{abstract}
\Gist\/ is a tool that (a) solves the qualitative analysis problem of
turn-based probabilistic games with $\omega$-regular objectives; and (b)
synthesizes reasonable environment assumptions for synthesis of
unrealizable specifications. Our tool provides the first and efficient 
implementations of several reduction-based techniques to solve turn-based 
probabilistic games, and uses the analysis of turn-based probabilistic games 
for synthesizing environment assumptions for unrealizable specifications.
\end{abstract}

\section{Introduction}
\Gist\/ (Game solver from IST) is a tool for (a)~qualitative analysis of \emph{turn-based probabilistic
games ($2\half$-player games)} with $\omega$-regular objectives, and 
(b)~computing environment assumptions for synthesis of unrealizable 
specifications. 
The class of $2\half$-player games arise in several important applications 
related to verification and synthesis of reactive systems. Some key 
applications are: (a) synthesis of stochastic reactive systems; 
(b) verification of probabilistic systems; and (c) synthesis of unrealizable
specifications. We believe that our tool will be useful for the above
applications.

\smallskip\noindent{\bf $2\half$-player games.} 
$2\half$-player games are played on a graph by two players along with 
probabilistic transitions.
We consider $\omega$-regular objectives over infinite paths specified by 
parity, Rabin and Streett (strong fairness) conditions that can express 
all $\omega$-regular properties such as safety, reachability, liveness, 
fairness, and most properties commonly used in verification. 
Given a game and an objective, our tool determines
whether the first player has a strategy to ensure that the objective 
is satisfied with probability~1, and if so, it constructs 
such a witness strategy. Our tool provides the first implementation of 
qualitative analysis (probability~1 winning) of $2\half$-player games 
with $\omega$-regular objectives.

\smallskip\noindent{\bf Synthesis of environment assumptions.}
The synthesis problem asks to construct a finite-state reactive 
system from an $\omega$-regular specification. In practice, initial specifications 
are often unrealizable, which means that there is no system that implements 
the specification. A common reason for unrealizability is that assumptions 
on the environment of the system are incomplete. The problem of 
correcting an unrealizable specification $\Psi$ by computing an environment 
assumption $\Phi$ such that the new specification $\Phi \to \Psi$ 
is realizable was studied in~\cite{CHJ08}.
The work~\cite{CHJ08} constructs an assumption $\Phi$ that constrains only 
the environment and is as weak as possible. 
Our tool implements the algorithms of~\cite{CHJ08}.
We believe our implementation will be useful in analysis 
of realizability of specifications and computation of 
assumptions for unrealizable specifications.
%realizability of specifications.

\section{Definitions}\label{section:definition}
We first present the basic definitions of games and objectives.

\smallskip\noindent{\bf Game graphs.} 
A \emph{turn-based probabilistic game graph} (\emph{$2\half$-player game graph})
$\gamegraph =((S, E), (\SA,\SB,\SR),\trans)$ consists of a directed graph 
$(S,E)$, a partition $(\SA$, $\SB$,$\SR)$ of the finite set $S$ of states, 
and a probabilistic transition function $\trans$: $\SR \rightarrow \distr(S)$, 
where $\distr(S)$ denotes the set of probability distributions over the 
state space~$S$. 
The states in $\SA$ are the {\em player-$\PA$\/} states, where player~$\PA$
decides the successor state; the states in $\SB$ are the {\em 
player-$\PB$\/} states, where player~$\PB$ decides the successor state; 
and the states in $\SR$ are the {\em probabilistic\/} states, where
the successor state is chosen according to the probabilistic transition
function~$\trans$. 
We assume that for $s \in \SR$ and $t \in S$, we have $(s,t) \in E$ 
iff $\trans(s)(t) > 0$. %%and we often write $\trans(s,t)$ for $\trans(s)(t)$. 
%For technical convenience we assume that every state in the graph 
%$(S,E)$ has at least one outgoing edge.
%%For a state $s\in S$, we write $E(s)$ to denote the set 
%%$\set{t \in S \mid (s,t) \in E}$ of possible successors.
%
The {\em turn-based deterministic game graphs} (\emph{2-player game graphs})
are the special case of the $2\half$-player game graphs with $\SR = \emptyset$.

\smallskip\noindent{\bf Objectives.} We consider the three canonical
forms of $\omega$-regular objectives: Streett and its dual Rabin
objectives; and parity objectives.
The Streett objective consists of $d$ request-response pairs
$\set{(Q_1,R_1),(Q_2,R_2),\ldots, (Q_d,R_d)}$ where $Q_i$ denotes a
request and
$R_i$ denotes the corresponding response (each $Q_i$ and $R_i$ are subsets of 
the state space). The objective requires that if a request $Q_i$ happens 
infinitely often, then the corresponding response must happen infinitely often.
The Rabin objective is its dual. 
The parity (or Rabin-chain objective) is the special case of Streett objectives
when the set of request-responses 
$Q_1 \subset R_1 \subset Q_2 \subset R_2 \subset  Q_3 \subset \cdots \subset 
Q_d \subset R_d$ form a chain.

\smallskip\noindent{\bf Qualitative analysis.} The qualitative analysis for 
$2\half$-player games is as follows: the input is a 
$2\half$-player game graph, and an objective $\Phi$ (Streett, Rabin or parity 
objective), and the output is the set of states such that player~0 
can ensure $\Phi$ with probability~1.
For detailed description of game graphs, plays, strategies, objectives and 
notion of winning see~\cite{KrishThesis}.
We focus on qualitative analysis because:
a)~In applications like synthesis the relevant analysis is qualitative
analysis: the goal is to synthesize a system that behaves correctly
with probability~1; (b)~Qualitative analysis for probabilistic games is independent of 
the precise probabilities, and thus robust with imprecision in the 
exact probabilities (hence resilient to probabilistic modeling errors). 
The qualitative analysis can be done with discrete graph theoretic 
algorithms. %%, as compared to quantitative analysis that 
%%requires cumbersome numerical computation. 
Thus qualitative analysis is more robust and efficient, and our 
tools implements these efficient algorithms.

\section{Tool Implementation} 
Our tool presents a solution of the following two problems.

\smallskip\noindent{\bf Qualitative analysis of $2\half$-player games.} 
Our tool presents the first implementation for the qualitative 
analysis of $2\half$-player games with Streett, Rabin and parity objectives.
We have implemented the linear-time reduction for qualitative analysis of 
$2\half$-player Rabin and Streett games to $2$-player Rabin and Streett games 
of~\cite{CdAH05}, and the linear-time reduction for $2\half$-player 
parity games to $2$-player parity games of~\cite{CJH04}.
The $2$-player Rabin and Streett games are solved by reducing them to the
$2$-player parity games using the LAR (latest appearance records) 
construction~\cite{GH82}. The $2$-player parity games are solved using the
tool PGSolver~\cite{Lange09}. 

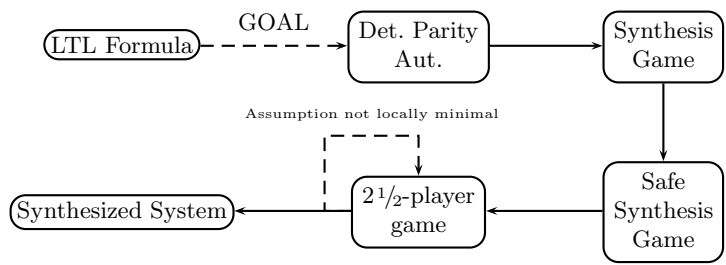
\begin{figure}
\begin{center}
\begin{pspicture}[showgrid=false](-1.5,0)(10,2.7)
\begin{psmatrix}[rowsep=1.0,colsep=1.5]
\psframebox[linearc=0.2,cornersize=absolute,framesep=2pt]{LTL Formula} & 
\psframebox[linearc=0.2,cornersize=absolute,framesep=2pt]{\tabular{c} Det. Parity \\ Aut. \endtabular} & 
\psframebox[linearc=0.2,cornersize=absolute,framesep=2pt]{\tabular{c} Synthesis \\ Game \endtabular} \\
\psframebox[linearc=0.2,cornersize=absolute,framesep=2pt]{Synthesized System}  &
\psframebox[linearc=0.2,cornersize=absolute,framesep=2pt]{\tabular{c} $2\half$-player \\ game \endtabular} &
\psframebox[linearc=0.2,cornersize=absolute,framesep=2pt]{\tabular{c} Safe \\ Synthesis \\ Game \endtabular} \\
\end{psmatrix}
\ncline[linestyle=dashed]{->}{1,1}{1,2}\naput{GOAL} \ncline{->}{1,2}{1,3}
\ncline{->}{1,3}{2,3} \ncline{->}{2,3}{2,2} \ncline{->}{2,2}{2,1}
\ncloop[angleA=180,angleB=-270,loopsize=-1,linestyle=dashed]{->}{2,2}{2,2}\naput{\tiny{Assumption not locally minimal}}
\end{pspicture}
\end{center}
\caption{An example illustrating the flow of the tool}
\label{fig:example}
\end{figure}

\smallskip\noindent{\bf Environment assumptions for synthesis.} 
Our tool implements a two-step algorithm for computing the environment assumptions
as presented in~\cite{CHJ08}. 
The algorithm operates on the game graph that is used to answer the 
realizability question. 
First, a safety assumption that removes a minimal set of 
environment edges from the graph is computed. 
Second, a fairness assumption that puts fairness conditions on some of the
remaining environment edges is computed. 
The problem of finding a minimal set of fair edges is 
computationally hard~\cite{CHJ08}, and a reduction to $2\half$-player games 
was presented in~\cite{CHJ08} to compute a locally minimal fairness assumption.
The details of the implementation are as follows: given an LTL formula $\phi$,
the conversion to an equivalent deterministic
parity automaton is achieved through GOAL~\cite{Goal}. Our tool then converts
the parity automaton into a $2$-player parity game by splitting the states and
transitions based on input and output symbols. Our tool then computes the safety
assumption by solving a safety model-checking problem. 
The computation of the fairness assumption is achieved in the following steps:
\begin{compactitem}
\item Convert the parity game with fairness assumption
into a $2\half$-player game.
\item Solve the $2 \half$-player game (using our tool) to check whether the
assumption is sufficient (if so, go to the previous step with a weaker fairness assumption).
\end{compactitem}
The synthesized system is obtained from a witness strategy of the parity
game. 
The flow is illustrated in Figure~\ref{fig:example}.

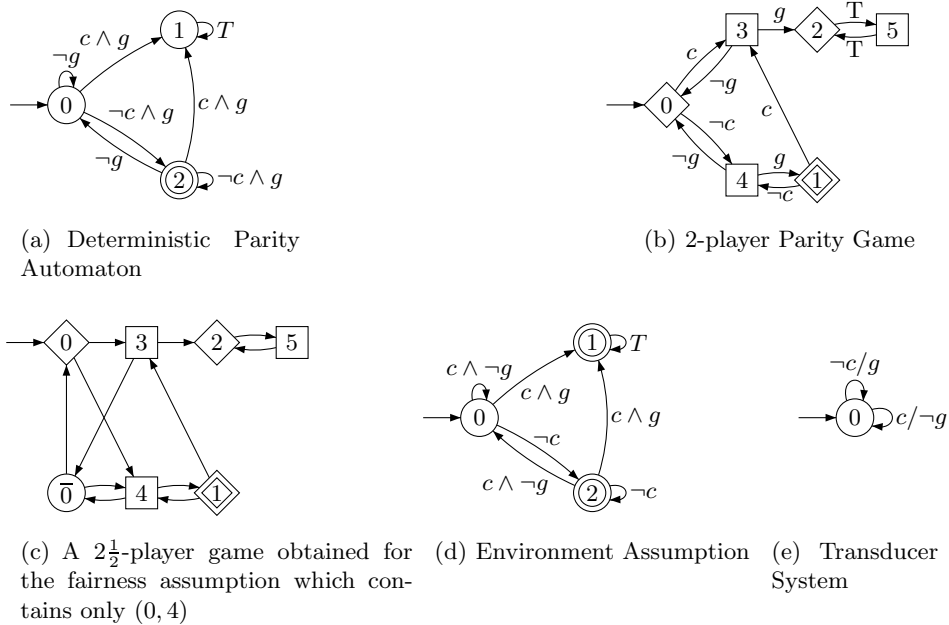
\begin{figure}[h]
\subfigure[Deterministic Parity Automaton]{
\begin{picture}(35,30)(0,0)
	\node[Nmarks=i,Nw=5,Nh=5,Nmr=2.5](0)(5,15){0}
	\node[Nw=5,Nh=5,Nmr=2.5](1)(20,25){1}
	\node[Nw=5,Nh=5,Nmr=2.5,Nmarks=r](2)(20,5){2}
	\drawloop[loopdiam=2,ELdist=0.5](0){$\neg g$}
	\drawedge[curvedepth=1,ELdist=0.5](0,1){$c \wedge g$}
	\drawedge[curvedepth=1,ELdist=0](0,2){$\neg c \wedge g$}
	\drawloop[loopdiam=2,ELdist=0.5,loopangle=0](1){$T$}
	\drawedge[curvedepth=1,ELdist=0.5](2,0){$\neg g$}
	\drawedge[curvedepth=-2,ELdist=0.5,ELside=r](2,1){$c \wedge g$}
	\drawloop[loopdiam=2,ELdist=0.5,loopangle=0](2){$\neg c \wedge g$}
\end{picture}
\label{fig:parity_automaton}
}
\subfigure[$2$-player Parity Game]{
\begin{picture}(40,30)(0,0)
	\rpnode[Nmarks=i,fangle=45](0)(5,15)(4,3){0}
	\rpnode[fangle=45,Nmarks=r](1)(25,5)(4,3){1}
	\rpnode[fangle=45](2)(25,25)(4,3){2}
	\rpnode[polyangle=45](3)(15,25)(4,3){3}
	\rpnode[polyangle=45](4)(15,5)(4,3){4}
	\rpnode[polyangle=45](5)(35,25)(4,3){5}

	\drawedge[curvedepth=1,ELdist=0.5](0,3){$c$}
	\drawedge[curvedepth=1,ELdist=0.5](3,0){$\neg g$}
	\drawedge[curvedepth=1,ELdist=0.5](0,4){$\neg c$}
	\drawedge[curvedepth=1,ELdist=0.5](4,0){$\neg g$}
	\drawedge(3,2){$g$}
	\drawedge(1,3){$c$}
	\drawedge[curvedepth=1,ELdist=0.5](4,1){$g$}
	\drawedge[curvedepth=1,ELdist=0.5](1,4){$\neg c$}
	\drawedge[curvedepth=1,ELdist=0.5](5,2){$\mbox{T}$}
	\drawedge[curvedepth=1,ELdist=0.5](2,5){$\mbox{T}$}
\end{picture}
\label{fig:synthesis_game}
}
\subfigure[A $2\frac{1}{2}$-player game obtained for the fairness
	assumption which contains only $(0,4)$]{
\begin{picture}(50,30)(0,0)
	\rpnode[Nmarks=i,fangle=45](0)(5,25)(4,3){0}
	\node[Nw=5,Nh=5,Nmr=2.5](0d)(5,5){$\overline{0}$}
	\rpnode[fangle=45,Nmarks=r](1)(25,5)(4,3){1}
	\rpnode[fangle=45](2)(25,25)(4,3){2}
	\rpnode[polyangle=45](3)(15,25)(4,3){3}
	\rpnode[polyangle=45](4)(15,5)(4,3){4}
	\rpnode[polyangle=45](5)(35,25)(4,3){5}

	\drawedge(0,3){}
	\drawedge[curvedepth=1,ELdist=0.5](0d,4){}
	\drawedge(0d,0){}
	\drawedge(3,0d){}
	\drawedge(0,4){}
	\drawedge[curvedepth=1,ELdist=0.5](4,0d){}
	\drawedge(3,2){}
	\drawedge(1,3){}
	\drawedge[curvedepth=1,ELdist=0.5](4,1){}
	\drawedge[curvedepth=1,ELdist=0.5](1,4){}
	\drawedge[curvedepth=1,ELdist=0.5](5,2){}
	\drawedge[curvedepth=1,ELdist=0.5](2,5){}
\end{picture}
\label{fig:probabilistic_game}
}\hfill
\subfigure[Environment Assumption]{
\begin{picture}(40,30)(0,0)
	\node[Nmarks=i,Nw=5,Nh=5,Nmr=2.5](0)(5,15){0}
	\node[Nw=5,Nh=5,Nmr=2.5,Nmarks=r](1)(20,25){1}
	\node[Nw=5,Nh=5,Nmr=2.5,Nmarks=r](2)(20,5){2}
	\drawloop[loopdiam=2,ELdist=0.5](0){$c \wedge \neg g$}
	\drawedge[curvedepth=1,ELdist=0.5,ELside=r](0,1){$c \wedge g$}
	\drawedge[curvedepth=1,ELdist=0](0,2){$\neg c$}
	\drawloop[loopdiam=2,ELdist=0.5,loopangle=0](1){$T$}
	\drawedge[curvedepth=1,ELdist=0.5](2,0){$c \wedge \neg g$}
	\drawedge[curvedepth=-2,ELdist=0.5,ELside=r](2,1){$c \wedge g$}
	\drawloop[loopdiam=2,ELdist=0.5,loopangle=0](2){$\neg c$}
\end{picture}
\label{fig:environment_assumption}
}\hfill
\subfigure[Transducer System]{
\begin{picture}(20,30)(0,0)
	\node[Nmarks=i,Nw=5,Nh=5,Nmr=2.5](0)(10,15){0}
	\drawloop[loopdiam=2.5,ELdist=0.5](0){$\neg c / g$}
	\drawloop[loopdiam=2.5,ELdist=0.5,loopangle=0](0){$c / \neg g$}
\end{picture}
\label{fig:transducer_system}
}
\caption{An example that illustrates the tool flow}
\label{fig:worked_example}
\end{figure}

We illustrate the working of our tool on a simple example shown in
Figure~\ref{fig:worked_example}
Consider an LTL formula $\Phi=GF \mathtt{grant} \wedge G(\mathtt{cancel} \to \neg 
\mathtt{grant})$, where $G$ and $F$ denote globally and eventually, respectively.
The propositions \texttt{grant} and \texttt{cancel} are abbreviated as \texttt{g} 
and \texttt{c}, respectively. 
From $\Phi$ our tool obtains a deterministic parity automaton (Figure~\ref{fig:parity_automaton})
that accepts exactly the words that satisfies $\Phi$. 
The parity automaton is then converted into a parity game. In Figure~\ref{fig:synthesis_game},
$\Box$ represents player~0 states and $\Diamond$ represents player~1 states. 
It can be shown that in this game no safety assumption required.
%% as there is no way the environment can force the play outside the cooperative winning region for the system's objective.
We illustrate how to compute a locally minimal fairness assumption. 
Given an fairness assumption on edges, our tool reduces the game with the assumption to a $2\half$-player parity game
(see details in~\cite{CHJ08}). If the initial state in the $2\half$-player game is in winning with probability~1 
for player~0, then the assumption is sufficient. Figure~\ref{fig:probabilistic_game} illustrates 
the $2\half$-player game obtained with the fairness assumption on the edge
$(0,4)$. The $\bigcirc$ state is 
the probabilistic state with uniform distribution over its successors.
The assumption on this edge is the minimal fairness assumption for the
example. Our tool then converts this game back into an automaton to
obtain the environment assumption as an
automaton(Figure~\ref{fig:environment_assumption}). This assumption
is equivalent to the formula $G(\neg (\mathtt{cancel \wedge grant})) \implies GF(\mathtt{\neg cancel})$. From a witness
strategy in Figure~\ref{fig:probabilistic_game}
our tool obtains the system that implements the specification with the
assumption (Figure~\ref{fig:transducer_system}).

\smallskip\noindent{\bf Performance of \Gist.} Our implementation of reduction 
of $2\half$-player games to $2$-player games is linear time and efficient, and
the computationally expensive step is solving $2$-player games. 
For qualitative analysis of $2\half$-player games, \Gist\ can handle game 
graphs of size that can be typically handled by tools solving $2$-player games.
Typical run-times for qualitative analysis of $2\half$-player parity
games of various sizes are summarized in Table~\ref{table:runtimes}. The
games used were generated using the benchmark tools of PGSolver and then
converting one-tenth of the states into probabilistic states. 

\begin{table}[ht]
\begin{center}
\label{table:runtimes}
\begin{tabular}{|c|c||c|c|c|}
%\begin{tabular}{|p{0.6in}|p{0.6in}||p{0.6in}|p{0.6in}|p{0.6in}|}
\hline
States & Edges & \multicolumn{3}{|c|}{Runtime (sec.)} \\
\cline{3-5}
&& Avg. & Best & Worst \\
\hline
1000 & 5000 & 1.17 & 0.63 & 1.59 \\
5000 & 25000 & 15.94 & 11.10 & 19.46 \\
10000 & 50000 & 51.43 & 39.38 & 62.61 \\
20000 & 100000 & 282.24 & 267.40 & 310.11 \\
50000 & 250000 & 2513.18 & 2063.40 & 2711.23 \\
\hline
\end{tabular}
\end{center}
\caption{Runtimes for solving $2\half$-player parity games}
\end{table}
In the case of synthesis of environment assumptions, the expensive step is the
reduction of LTL formula to deterministic parity automata. Our tool can 
handle formulas that are handled by classical tools for translation of LTL 
formula to deterministic parity automata.

\smallskip\noindent{\bf Other features of \Gist.}
Our tool is compatible with several other game solving and synthesis tools: 
\Gist\/  is compatible with PGSolver and GOAL. Our tool provides a graphical 
interface to describe games and automata, and thus can also be used as a 
front-end to PGSolver to graphically describe games.

%%\bibliographystyle{plain}
%%\bibliography{main}

\newpage
\section{Appendix: Details of the Tool}
\Gist~is available for download at \GistUrl~for Unix-based architectures. All
the libraries that \Gist~uses are packaged along with it.

\subsection{Dependencies and Architecture}
\smallskip\noindent{\bf Language, tools and installation.}
\Gist~is written in Scala and it uses several other tools. 
For the graphical interface to draw game graphs and automata it uses the 
JUNG library~\cite{website:jung} for graph layout algorithms.
For translation of an LTL formula to a deterministic parity automata it uses 
GOAL~\cite{Goal}.
The solution of $2$-player parity games is achieved by using PGSolver~\cite{Lange09}. 
%%\par
For compilation and installation: (a) an installation of the Scala compiler and
runtime environment is required; (b) the PGSolver build process requires an OCaml
compiler to be installed; and (c) GOAL and JUNG require a Java runtime environment to
be installed.

\smallskip\noindent{\bf Source code.}
The source code of \Gist~is composed mainly of five modules:
\begin{enumerate}
\item Module \textbf{newgames}  mainly consists of the classes for
probabilistic $\omega$-regular games, i.e.\ games with B\"{u}chi, coB\"{u}chi, Rabin, Streett
and parity objectives. 
Each of these classes contains routines for the reduction of the $2\half$-player version
to the $2$-player version which preserves the probability 1 winning
region of Player 1.
Each of these classes also returns a witness strategy for the player as required.

\item Module \textbf{specification} consists of classes implementing the
specifications for the synthesis problem, i.e. LTL formulae, B\"{u}chi automata
and parity automata. The class for LTL formulae contains a routine to convert
LTL formulae into an equivalent nondeterministic B\"{u}chi automata and the class 
for B\"{u}chi automata has a routine for converting it into a deterministic parity automaton. 
The parity automata class can generate the synthesis game (by splitting transitions) 
for the automaton as described in~\cite{CHJ08}.

\item Module \textbf{synthesis}  contains the classes relevant to the process of
synthesis. The class for synthesis games contains routines (a)~to compute
transducers implementing the specification; (b)~to compute minimal safety assumption
and locally minimal fairness assumption in case of an unrealizable
specification; (c)~to check whether user-specified assumptions are sufficient to
make the specification realizable; and (d)~to get the assumptions as a Streett
automaton. 
\item Modules \textbf{gui} and \textbf{cui} contain classes for graphical and
text based user interfaces. Most of the functionality in the \textbf{cui} module
is contained in the \textbf{Console} class, which interprets a command
line. The \textbf{gui} module contains forms and windows to display
various automata and games; and provide an interface for the various
operations on them.
\item Module \textbf{basic} contains the definitions which are needed by all
other packages, namely, the classes for alphabet, symbols and generic automata.
\end{enumerate}
In addition to these, there are other routines to parse and write automata and game 
graphs in files in a format that can be used with GOAL.

\subsection{User Manual}

In this section we describe the usage of the graphical and text-based interface
of the tool. \\
{\bf Format of files.} The file format used by the tool is based
on the format used by GOAL. The format for games and automata structures is
presented below:

\begin{verbatim}
<structure label-on="transition" type=["game"|"fa"]>
  <alphabet type="propositional">
    <prop type=["input"|"output"]>TEXT</prop>
    ...
  </alphabet>
  <stateSet>
    <state sid="NUMERIC">
      [<player>[0|1|-1]</player>]
      [<label>TEXT</label>]
    </state>
    ...
  </stateSet>
  <transitionSet>
    <transition tid="NUMERIC">
      <from>NUMERIC(State ID)</from>
      <to>NUMERIC(State ID)</to>
      <read>TEXT(Symbol)</read>
    </transition>
    ...
  </transitionSet>
  <initialStateSet>
    <stateID>NUMERIC</stateID>
  </initialStateSet>
  <acc type="[buchi|parity|rabin|streett]">
    <accSet> %Ony one set for Buchi acceptance condition.
      <stateID>NUMERIC(State ID)</stateID>
      ...
    </accSet>
    <accSet> %Multiple sets for Parity acceptance condition.
             %One for each priority
      <stateID>NUMERIC(State ID)</stateID>
      ...
    </accSet>
    <accSet> %Multiple sets for Rabin and Streett acceptance conditions.
             %Different format from the other conditions.
      <E>
        <stateID>NUMERIC(State ID)</stateID>
        ...
      </E>
      <F>
        <stateID>NUMERIC(State ID)</stateID>
        ...
      </F>
    ...
    </accSet>
  </acc>
</structure>
\end{verbatim}

\noindent{\bf Graphical Interface.} The graphical interface for \Gist~consists
of a window for each kind of game graph, automata, and formula the tool
handles.
When \Gist\/ is invoked, a window is shown with buttons for each kind.
A window for a specific kind contains buttons that represent relevant actions
that can be performed.
There are also generic options such as saving and loading.

For automata and game graphs, the window contains an area in which the graph
is laid out visually. The layout can be changed by dragging the vertices and the
edges of the graph.  \Gist~uses the layout algorithms of JUNG to
automatically layout the graph. The layout algorithms can be chosen by right-clicking 
on the window and selecting \textbf{Layout} from a pop-up menu that appears.
Also, sets of vertices or edges can be highlighted for other operations (such as
finding sufficiency of assumptions containing these edges) by choosing
\textbf{Highlight Mode} on the pop-up menu.

The tool also includes interfaces for building automata and games graphically.
In these windows, one can insert states or edges into a structure by selecting
the appropriate mode from the pop-up menu. When an edge is created, the user can
label the edge appropriately. The alphabet for the symbols (for labeling edges) must be
set before the edges are created. States and edges can also be deleted
using the \textbf{Delete mode}.

\noindent{\bf Text-based Interface.} The text-based interface for \Gist~is an
interactive prompt. The user can define and use variable for any object.
Variables need not be declared before use. All variable names need to begin with
a \$. The syntax for the statements is defined below.
\begin{verbatim}
Variable := $[a-zA-Z0-9]*
Statement :=   Variable               --Prints the value of the variable
             | Variable = Variable    --Assignment
             | Variable = Expression  --Assignment
Expression := Object Action
Object :=   "LTL" | "BuchiAutomaton" | "ParityAutomaton" 
              | "SynthesisGame" | "StreettAutomaton" | "ParityGame" 
              | "RabinGame" | "StreettGame"
Action := readFile ... | writeFile ... | help | ...
\end{verbatim}
The ``action" as seen in the above syntax definition varies depending upon the
object. The \textbf{help} action for any object displays all the other actions
available for this object along with an explanation. 
\par
All objects which represent
games have the following actions: \textbf{winningRegion},
\textbf{cooperativeWinningRegion}, and \textbf{toDeterministicGame}. The action
\textbf{winningRegion} takes an argument, either 0 or 1 (for a player), and computes the
set of states from which the player wins with probability~1. The action
\textbf{cooperativeWinningRegion} is invoked only for $2$-player games, and 
it computes the set of states such that there is a path to satisfy the 
objective of player~0.
The action \textbf{toDeterministicGame} is invoked on $2\half$-player games and it
returns a $2$-player game in which probability~1 winning of player 0 is
preserved. 
In addition, the action \textbf{winningStrategy} computes the winning strategy of 
each player in $2$-player games, and the probability~1 winning strategy in $2\half$-player 
games.

The objects for B\"{u}chi automata have an action \textbf{toParityAutomaton} to
convert it into equivalent deterministic parity automata. Similarly, the objects 
for LTL formulae and parity automata have actions to convert them into nondeterministic 
B\"{u}chi automata and Synthesis games respectively. The objects for Synthesis games 
have actions related to synthesis and computation of environment assumptions.

\begin{figure}
\begin{center}
\includegraphics[width=0.8\textwidth]{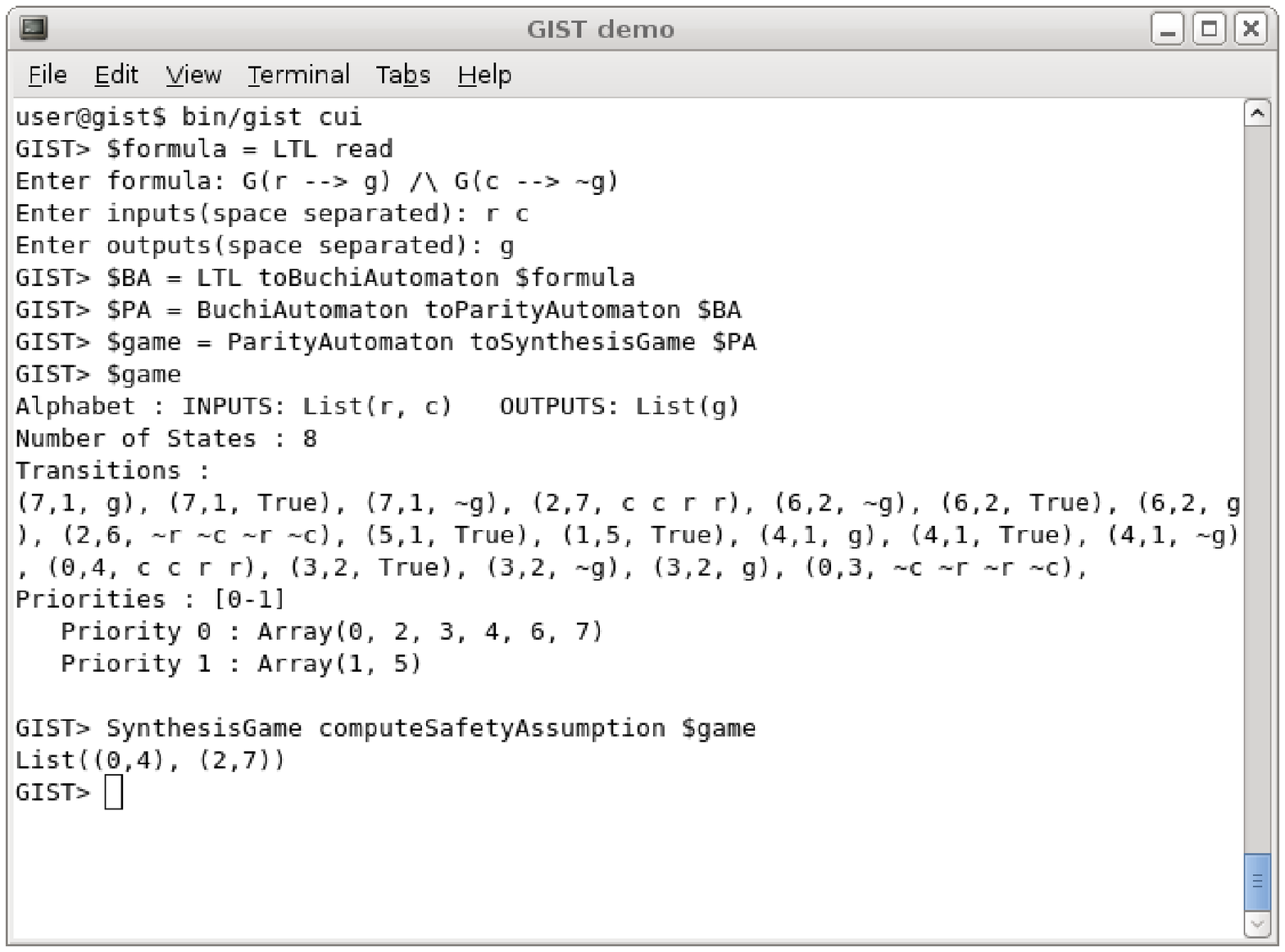}
\end{center}
\caption{Example to illustrate the text-based interface}
\label{fig:text_interface}
\end{figure}

\begin{figure}
\begin{center}
\includegraphics[width=0.8\textwidth]{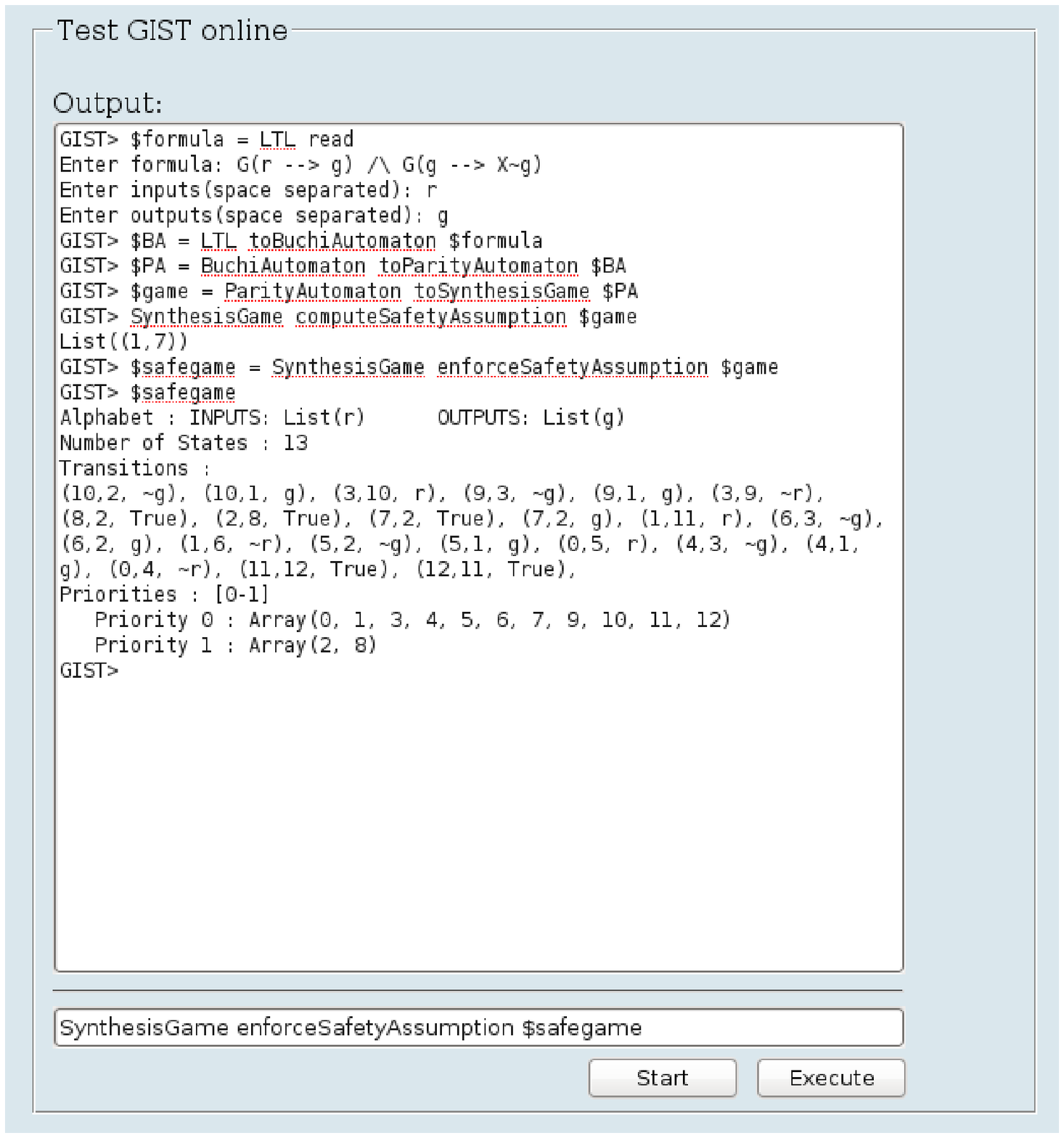}
\end{center}
\caption{\Gist~web interface}
\label{fig:web_interface}
\end{figure}

The text-based interface for \Gist~is also available online at \GistUrl.
Figure~\ref{fig:text_interface} shows the screenshot for the text-interface with 
input and output for Example~1 (described in the following subsection).
Figure~\ref{fig:web_interface} shows the screenshot of the web interface for a 
similar example.

\subsection{Examples to Illustrate the Usage of \Gist}

In this section, we present two examples to illustrate  the usage of \Gist. 
We have chosen small examples for the simplicity of the presentation to 
illustrate  the usage of \Gist.
These examples demonstrate the usage of \Gist~for computation of 
environment assumptions for synthesis and uses solution of 
$2\half$-player games. 
In these examples, we compute the assumptions for two unrealizable specifications given in 
LTL. Both the specifications are about request-response
systems and are chosen to illustrate safety and fairness assumptions respectively.
%%The examples are in the same vein as ones from \cite{CHJ08}.

\begin{figure}[htb]
\includegraphics[width=\textwidth]{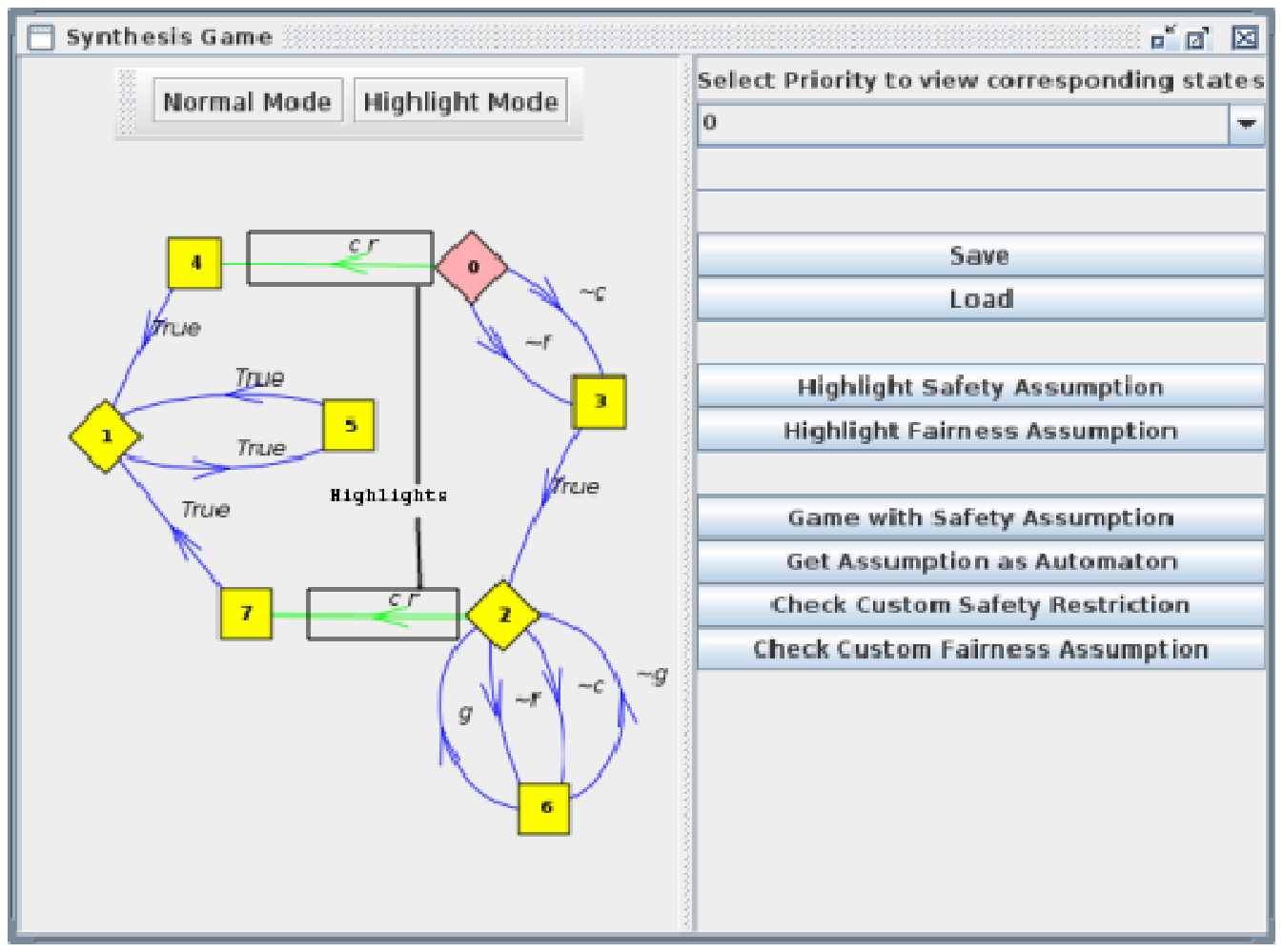}
\caption{Example 1. The safety assumption is highlighted}
\label{fig:safety_assumption}
\end{figure}
\emph{Example 1.} Consider a request-response system in which there are two
inputs, \texttt{request} and \texttt{cancel}, and one output \texttt{grant}.
Now, consider the specification $G (\mathtt{request} \to \mathtt{grant}) \wedge
G (\mathtt{cancel} \to \neg \mathtt{grant})$. This specification is unrealizable: 
any input in which both \texttt{request} and \texttt{cancel} are set at the same time
does not have an output which satisfies the specification.
We can compute an environment assumption for this specification using \Gist.
Intuitively, we would want an assumption that says \texttt{request} and
\texttt{cancel} must not be set at the same time provided the specification was not already
violated earlier. We show that the assumption can be computed automatically 
by \Gist.

To compute the assumption using \Gist, we select LTL formula from the main window of
options and then enter the formula above, specifying the inputs and outputs.
This formula is then converted into a nondeterministic B\"{u}chi automaton and then to 
a deterministic parity automaton, and finally to a synthesis game. 
In this game, we attempt to compute the safety assumption. The safety assumption is 
highlighted (green arrows in a box; (0,4) and (2,7)) as shown in Figure~\ref{fig:safety_assumption}.
As shown in Figure~\ref{fig:safety_assumption}, the 
safety assumption includes all the edges where \texttt{request} and \texttt{cancel} 
are set at the same time. But, if there has been an instance of a \texttt{request} not being
granted already, then there is no restriction on the inputs. This is the same
assumption as was expected intuitively. Now, we can obtain a synthesis game
where the safety assumption is enforced. In this new game, if the fairness assumption is computed
the output shows no fairness assumption is necessary. A transducer that
implements the modified specification can be obtained from the solution of this game.

\begin{figure}[htb]
\begin{center}
\includegraphics[width=0.8\textwidth]{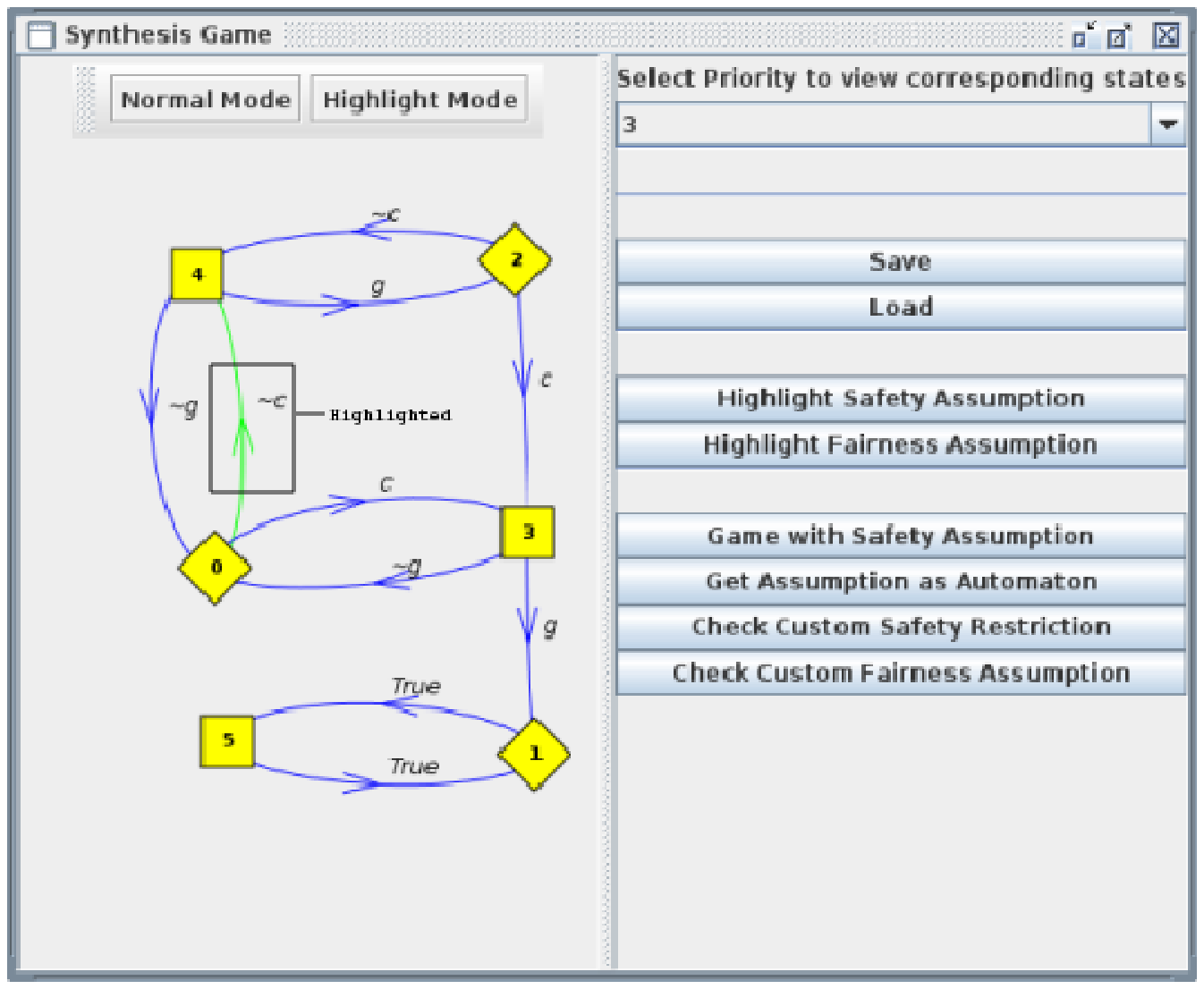}
\end{center}
\caption{Example 2. The fairness assumption is highlighted}
\label{fig:fairness_assumption}
\end{figure}
\emph{Example 2.} Consider the request-response system as in Example 1. But, with the 
specification $(G F \mathtt{grant}) \wedge G (\mathtt{cancel} \to \neg
\mathtt{grant})$. This specification says that we should have infinitely many
grants and that at every step, if \texttt{cancel} is set, then there should be
no grant at that step. This specification is also unrealizable as any input
where the \texttt{cancel} is always set has no acceptable output. We can
see that if \texttt{cancel} is not set always after a point, then the
specification becomes realizable. This condition can be computed using \Gist\/ following 
the same steps as in the above example: first the tool finds that 
no safety assumption is necessary, and then it computes the fairness assumption in the
synthesis game. The fairness assumption is computed internally by reduction 
to $2\half$-player games. The fairness assumption is highlighted (by green arrow in a box; (0,4))
in the screenshot Figure~\ref{fig:fairness_assumption}. 
The computed assumption can be interpreted as follows: the highlighted edge 
must be taken infinitely often if the source vertex of the edge is visited infinitely often. 
Translating this into propositions, it means that at any step, \texttt{cancel} cannot be 
set forever in the future. 

\end{document}